%% file: optimal_noc_for_nn.tex
\documentclass[acmsmall]{acmart}

\usepackage{booktabs} 

\usepackage{amsmath}
\usepackage{multirow}

\usepackage{multicol}
\usepackage{graphicx}
\usepackage[ruled, linesnumbered]{algorithm2e}
\usepackage[normalem]{ulem}

\usepackage{subcaption}
\setcopyright{rightsretained}

\newcommand{\JS}[1]{\textcolor{black}{#1}}
\newcommand{\rev}[1]{\textcolor{black}{#1}}
\newcommand{\minrev}[1]{\textcolor{black}{#1}}

\setcopyright{acmcopyright}
\acmJournal{JETC}
\acmYear{2021} \acmVolume{1} \acmNumber{1} \acmArticle{1} \acmMonth{1} \acmPrice{15.00}\acmDOI{10.1145/3460233}

\begin{document}
\title{Impact of On-Chip Interconnect on In-Memory Acceleration of Deep Neural Networks}

\thanks{*Authors have equal contributions.}

\author{Gokul Krishnan*}
\email{gkrish19@asu.edu}
\orcid{0000-0003-1085-2189}
\affiliation{%
	\institution{Arizona State University}
	\department{School of Electrical, Computer, and Energy Engineering}
	\city{Tempe}
	\state{AZ}
	\postcode{85287}
	\country{USA}
}

\author{Sumit K. Mandal*}
\email{skmandal@wisc.edu}
\orcid{0000-0003-1085-2189}
\affiliation{%
	\institution{University of Wisconsin-Madison}
	\department{Department of Electrical and Computer Engineering}
	\city{Madison}
	\state{WI}
	\postcode{53706}
	\country{USA}
}

\author{Chaitali Chakrabarti}
\email{chaitali@asu.edu}
\orcid{0000-0003-1085-2189}
\affiliation{%
	\institution{Arizona State University}
	\department{School of Electrical, Computer, and Energy Engineering}
	\city{Tempe}
	\state{AZ}
	\postcode{85287}
	\country{USA}
}

\author{Jae-sun Seo}
\email{jseo28@asu.edu}
\orcid{0000-0003-1085-2189}
\affiliation{%
	\institution{Arizona State University}
	\department{School of Electrical, Computer, and Energy Engineering}
	\city{Tempe}
	\state{AZ}
	\postcode{85287}
	\country{USA}
}

\author{Umit Y. Ogras}
\email{uogras@wisc.edu}
\affiliation{%
	\institution{University of Wisconsin-Madison}
	\department{Department of Electrical and Computer Engineering}
	\city{Madison}
	\state{WI}
	\postcode{53706}
	\country{USA}
}

\author{Yu Cao}
\email{Yu.Cao@asu.edu}
\orcid{0000-0003-1085-2189}
\affiliation{%
	\institution{Arizona State University}
	\department{School of Electrical, Computer, and Energy Engineering}
	\city{Tempe}
	\state{AZ}
	\postcode{85287}
	\country{USA}
}

\input{abstract.tex}




\setcopyright{acmcopyright}
\acmJournal{JETC}

\maketitle

\input{introduction.tex}

\input{relatedwork.tex}
\vspace{3mm}
\input{simulation}

\input{noc_performance}

\input{Interconnect.tex}
\input{expt_results.tex}
\input{conclusion.tex}

\input{ack.tex}
\bibliographystyle{ACM-Reference-Format}
\bibliography{ref}

\end{document}

%% file: abstract.tex
\begin{abstract}
With the widespread use of Deep Neural Networks (DNNs), machine learning algorithms have evolved in two diverse directions -- one with ever-increasing connection density for better accuracy
and the other with more compact sizing for energy efficiency. The increase in connection density increases on-chip data movement, which makes efficient on-chip communication a critical function of the DNN accelerator. The contribution of this work is threefold. 
First, we illustrate that the point-to-point (P2P)-based interconnect is incapable of handling a high volume of on-chip data movement for DNNs. Second, we evaluate P2P and network-on-chip (NoC) interconnect \minrev{(with regular topology)} for SRAM- and ReRAM-based in-memory computing (IMC) architectures for a range of DNNs. 
This analysis shows the necessity for the optimal interconnect choice for an IMC DNN accelerator. Finally, we perform an experimental evaluation for different DNNs to empirically obtain the performance of the IMC architecture with both NoC-tree and NoC-mesh.
We conclude that, at the tile-level, NoC-tree is appropriate for compact DNNs employed at the edge, and NoC-mesh is necessary to accelerate DNNs with high connection density.
Furthermore, we propose a technique to determine the optimal choice of interconnect for any given DNN.
In this technique, we use analytical models of NoC to evaluate end-to-end communication latency of any given DNN.
We demonstrate that the interconnect optimization in the IMC architecture results in up to 6$\times$ improvement in energy-delay-area product for VGG-19 inference compared to the state-of-the-art ReRAM-based IMC architectures.

\end{abstract}

%% file: introduction.tex
\section{Introduction}\label{sec:intro}


DNNs have achieved high accuracy that exceeds human-level perception
for a variety of applications such as computer vision, natural language processing, and medical imaging~\cite{krizhevsky2012imagenet, deng2013new, litjens2017survey}.
The DNNs that achieve higher accuracy tend to consist of deeper and denser network structures.
On the other hand, DNNs for edge devices tend to use smaller and shallower networks.

\begin{figure}[t]
	\centering
	\includegraphics[width=0.6\textwidth]{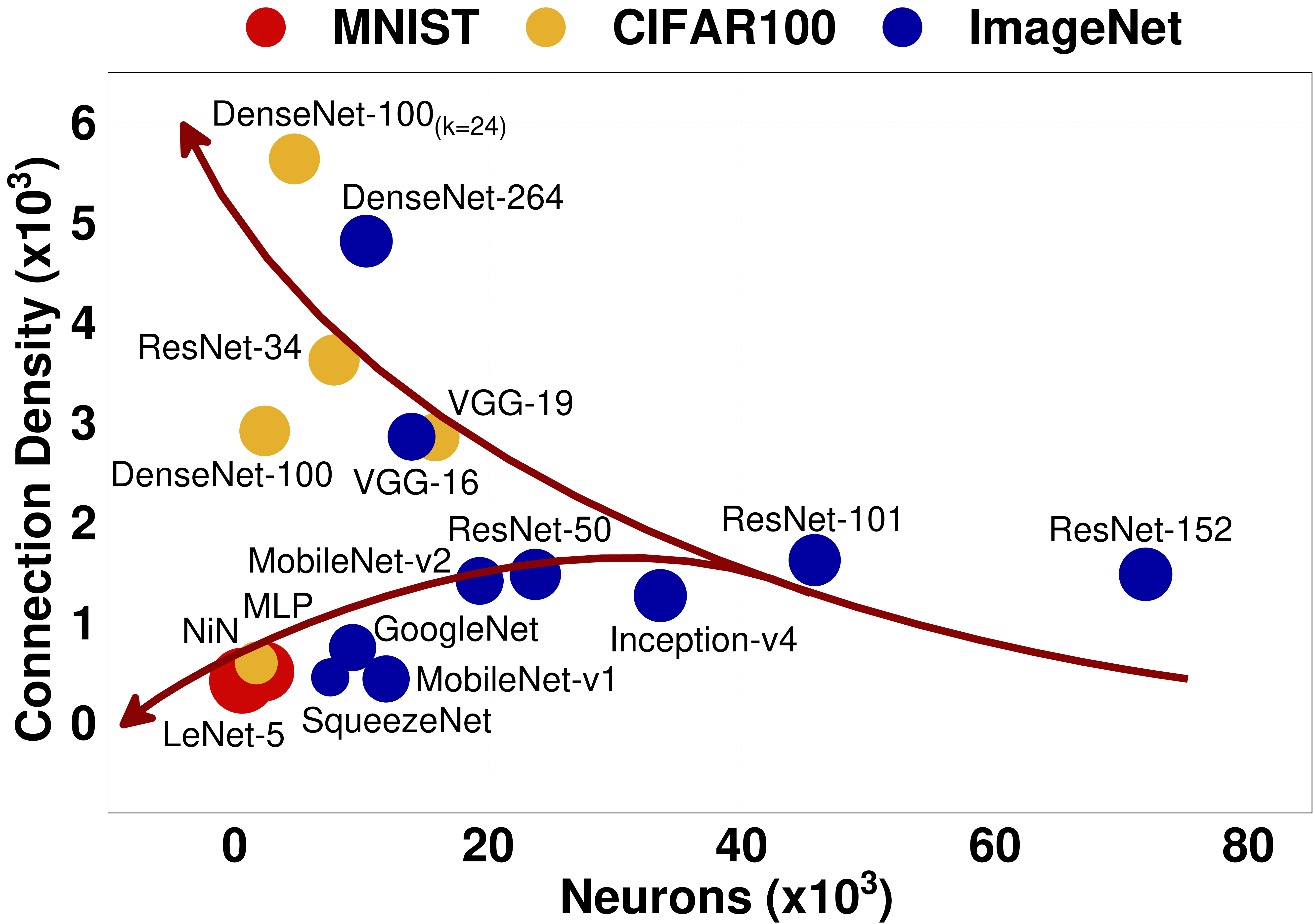}
	\caption{Connection density of different DNNs for three different datasets. Each output feature map (convolution layer) and neural unit (FC layer) represent a neuron. Larger markers represent higher accuracy.} 
	\label{fig:ANN}
 \end{figure}

\begin{figure}[h]
	\centering
	\includegraphics[width=0.7\textwidth]{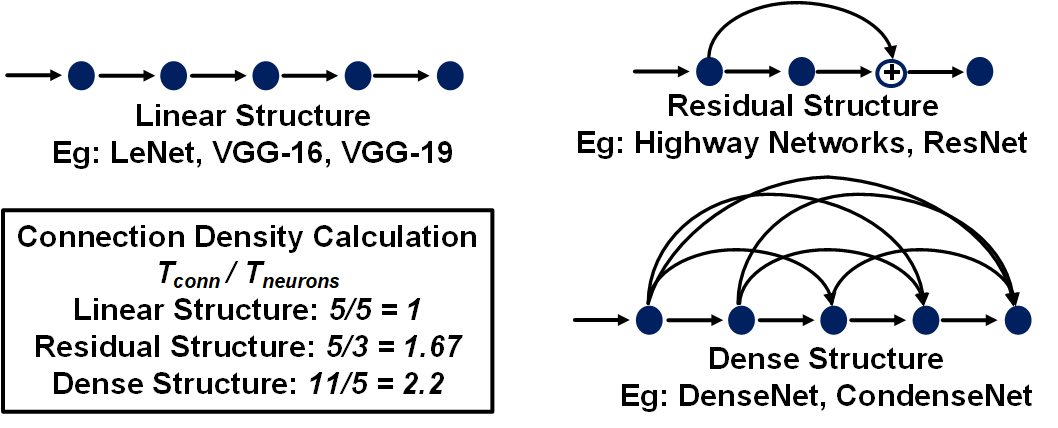}
	\caption{Different types of DNN structures and their representative connection density.}
	\label{fig:linear_nonlinear}
\end{figure}

Figure~\ref{fig:ANN} shows the trend in connection density for various DNNs in the literature,
where \textit{connection density} is defined as the average number of connections per neuron in DNNs. In the context of DNNs, a neuron is defined as an output feature of a convolution layer and every neural unit of the fully-connected (FC) layer.
Three representative DNN structures and connection patterns are illustrated in Figure~\ref{fig:linear_nonlinear}. 
Linear structures such as LeNet-5~\cite{lecun1998gradient} and VGG-19~\cite{simonyan2014very} have a connection density of one owing to one connection per neuron.
Since residual networks such as ResNet~\cite{he2016deep} have residual skips, it has more connections than the number of neurons resulting in a connection density higher than one. Dense structures like DenseNet~\cite{huang2017densely} have multiple connections from each neuron, resulting in a higher connection density.

We observe two main trends by analyzing the connection density for different DNNs in Figure~\ref{fig:ANN}.
First, increasing connection density provides higher accuracy, which is essential for cloud-based computing platforms.
Second, lower connection density is observed for compact models, which is necessary for edge computing hardware.
Both hardware platforms require the processing of large amounts of data with corresponding power and performance constraints.
Hence, there is a need to design optimal hardware architectures with low power and high performance for DNNs with different connection densities.


With limited on-chip memory, conventional DNN architectures inevitably involve a significant amount of communication with off-chip memory resulting in increased energy consumption~\cite{chen2019eyeriss}.
However, it has been reported that the energy consumption of off-chip communication is 1,000$\times$ higher than the energy required to perform the computations~\cite{horowitz20141}. 
Dense structures like DenseNet perform approximately $2.7\times10^7$ off-chip memory accesses to process a frame of an image~\cite{huang2017densely}. 
As a result, off-chip memory access becomes the energy bottleneck for hardware architectures of dense structures.
Employing dense embedded non-volatile memory (NVM) such as ReRAM for in-memory computing (IMC) substantially reduces off-chip memory accesses~\cite{shafiee2016isaac,song2017pipelayer}.
%
\begin{figure}[b]
	\centering
	\includegraphics[width=0.6\textwidth]{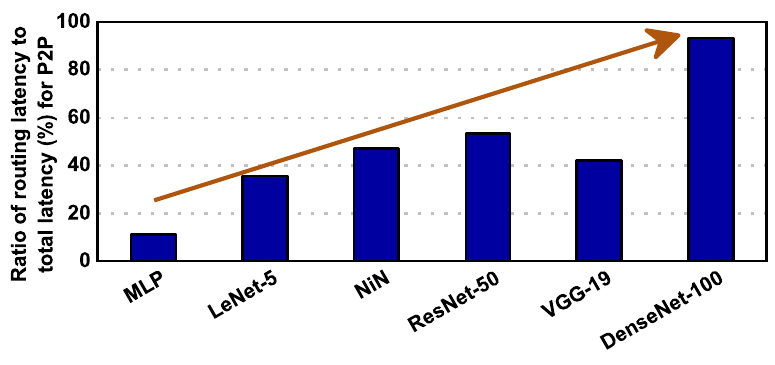}
	\caption{Contribution of routing latency to total latency for different DNNs for a P2P-based IMC architecture~\cite{chen2018neurosim}. With increase in connection density, routing latency becomes the bottleneck for performance.} 
	\label{fig:cnn}
\end{figure}

On-chip interconnect is an integral part of hardware architectures that incorporate in-memory acceleration.
Both point-to-point (P2P) interconnect~\cite{kwon2018maeri, venkataramani2017scaledeep} and NoC-based interconnect~\cite{shafiee2016isaac, chen2019eyeriss, krishnan2020interconnect} are used for on-chip communication in state-of-the-art DNN accelerators. 
Shafiee \textit{et al.}~\cite{shafiee2016isaac} utilizes a concentrated mesh for the interconnect, while Chen \textit{et al.}~\cite{chen2019eyeriss} employs three different NoCs that are used for on-chip data movement in the architecture. In contrast, Krishnan \textit{et al.}~\cite{krishnan2020interconnect} utilizes a custom mesh-NoC for on-chip communication. The custom NoC derives the structure based on the on-chip traffic between different IMC processing elements (PEs), where each PE denotes the SRAM- or ReRAM-based IMC crossbar.
\rev{A technique to construct custom NoC which provides minimum communication latency for a given DNN is proposed in~\cite{mandal2020latency}.
Since custom NoC requires alteration in hardware for different DNNs, our studies focus on regular NoC topologies.
A more detailed survey on work which design efficient interconnect for DNN accelerators can be found in~\cite{nabavinejad2020overview}.}

To better understand the need for an NoC-based on-chip interconnect, we analyze the scalability of P2P interconnect in in-memory computing (IMC) architectures by evaluating the contribution of routing latency to end-to-end latency for different DNNs, as shown in Figure~\ref{fig:cnn}.
The contribution of routing latency increases up to 94\% with increasing connection density. 
The high routing latency is attributed to the increased connection density, which correlates to more on-chip data movement. 
VGG-19 shows a reduced contribution compared to lower connection density DNNs due to the high utilization of the IMC PEs or crossbars resulting in reduced on-chip data movement.
Hence, P2P networks do not provide a scalable solution for high connection density DNNs. 
At the same time, NoC-based interconnects require higher area and energy for operation and can result in a significant overhead for low connection density DNNs.
Furthermore, different NoC topologies, mesh, or tree, are appropriate for DNNs with varying connection densities.
Therefore, a connection density-aware interconnect solution is critical to DNN acceleration.


In this work, we first perform an in-depth performance analysis of P2P interconnect-based in-memory computing (IMC) architectures~\cite{song2017pipelayer}. Through this analysis, we establish that P2P-based interconnects are incapable of handling data communication for dense DNNs and that NoC-based interconnect is needed for IMC architectures. 
Next, we evaluate P2P-based and NoC-based SRAM and ReRAM IMC architectures for a range of DNNs.
\rev{Further, we evaluate NoC-tree, NoC-mesh, and c-mesh topologies for the IMC architectures.
A c-mesh NoC is used in~\cite{shafiee2016isaac} at the tile-level to connect different tiles.
C-mesh uses more number of links and routers, providing better performance in terms of communication latency. 
However, interconnect area and energy becomes exorbitantly high for c-mesh NoC. 
Therefore, the energy-delay-area product (EDAP) of c-mesh is higher than NoC-mesh.
Hence, we restrict the detailed evaluations to NoC-mesh and NoC-tree.}
In these evaluations, we perform cycle-accurate NoC simulations through Booksim~\cite{jiang2013detailed}.
However, cycle-accurate NoC simulations are very time consuming and consequently slow down the overall performance analysis of IMC architectures.
Our experiment with different DNNs (the simulation framework is described in more detail in Section~\ref{sec:sim}) shows that cycle-accurate NoC simulation takes up to 80\% of the total simulation time for high connection density DNNs.
\begin{table}[t]
\caption{Summary of notations} \label{tab:notations}
\centering
\resizebox{0.6\textwidth}{!}{
\begin{tabular}{@{}ll|ll@{}}
\toprule
\textbf{Symbol} & \textbf{Definition}                                                                                & \textbf{Symbol}            & \textbf{Definition}                                                                                                                                                                   \\ \midrule
$N_L$  & \begin{tabular}[c]{@{}l@{}}Number of\\ Layers\end{tabular}                                  & $x_i$, $y_i$      & \begin{tabular}[c]{@{}l@{}}Input image size in \\ $i^{\textrm{th}}$ layer\end{tabular}                                                                                         \\ \midrule
$T_i$  & \begin{tabular}[c]{@{}l@{}}Number of \\ tiles in $i^{\textrm{th}}$ layer\end{tabular}       & $C_i$             & \begin{tabular}[c]{@{}l@{}}Input channels of \\ $i^{\textrm{th}}$ layer\end{tabular}                                                                                           \\ \midrule
$A_i$  & \begin{tabular}[c]{@{}l@{}}Number of \\ activations in $i^{\textrm{th}}$ layer\end{tabular} & $\lambda_{i,j,k}$ & \begin{tabular}[c]{@{}l@{}}Injection rate from $j^{\textrm{th}}$\\ tile of $(i-1)^{\textrm{th}}$ \\ layer to $k^{\textrm{th}}$ tile \\ of $i^{\textrm{th}}$ layer\end{tabular} \\ \midrule
$d_i$  & \begin{tabular}[c]{@{}l@{}}Input activations in \\ $i^{\textrm{th}}$ layer\end{tabular}     & $L_{comm}$     & Total communication latency                                                                                                                                                          \\ \bottomrule
\end{tabular}
}
\end{table}

To accelerate the overall performance analysis of the IMC architecture, we propose analytical models to estimate the NoC performance of a given DNN. Specifically,
we incorporate the analytical router modeling technique presented in~\cite{ogras2010analytical} to obtain the performance model for an NoC router.
Then we extend \rev{the existing} analytical model to get an estimation of end-to-end communication latency for NoC-tree and NoC-mesh for any given DNN as a function of the number of neurons and connection density.
\rev{Through the analytical latency model, the variable communication patterns of different DNNs are incorporated using connection density and number of neurons.}
Leveraging this analysis and the analytical model, we conclude the importance of the optimal choice of interconnect at different hierarchies of the IMC architecture. 
\rev{Utilizing the same analysis, we provide guidance for the optimal choice of interconnect for IMC architectures.}
At the tile-level, NoC-mesh for high connection density DNNs and an NoC-tree for low connection density DNNs provide low power and high performance for IMC-based architectures. 
Leveraging this observation, we propose an NoC-based heterogeneous interconnect IMC architecture for DNN acceleration. We demonstrate that the NoC-based heterogeneous interconnect IMC architecture (ReRAM) achieves up to 6$\times$ improvement in the energy-delay-area product (EDAP) for inference of VGG-19 when compared to state-of-the-art implementations.
The following are key contributions of this work: 
\begin{itemize}
    \item An in-depth analysis of the shortcomings of P2P-based interconnect and the need for NoC in IMC architectures.
    \item Analytical and empirical analysis to guide the choice of optimal NoC topology for an NoC-based heterogeneous interconnect.
    \item The proposed heterogeneous interconnect IMC architecture achieves 6$\times$ improvement in EDAP with respect to state-of-the-art ReRAM-based IMC accelerators.
\end{itemize}
%

The rest of the paper is organized as follows. Section~\ref{sec:background} introduces the background and motivation, Section~\ref{sec:sim} discusses the simulation framework used in this work, and Section~\ref{sec:ana_perf} presents the analytical performance modeling-based technique to obtain the optimal choice of NoC for any given DNN. Section~\ref{sec:arch} presents the in-memory architecture with heterogeneous interconnect. Section~\ref{sec:expt} discusses the experimental results, and Section~\ref{sec:conclusion} concludes the paper.

%% file: relatedwork.tex
\section{Motivation and Related Work}\label{sec:background}
\subsection{Deep Neural Networks}

We categorize DNNs into \JS{three} main classes, linear~\cite{simonyan2014very}, residual~\cite{he2016deep}, and dense~\cite{xie2019exploring}, as shown in Figure~\ref{fig:linear_nonlinear}. DNN structures include convolution layers stacked on top of each other for feature extraction and a set of classifier layers at the end to classify based on the features. 
A data point $d(x,y,c)$ in layer $i$+$1$ can be expressed using the notation summarized in Table~\ref{tab:notations} as follows:
\begin{equation}\label{eq:NN}
d_{i+1}(x,y,c) = \sum_{c_i=0}^{C_i-1} \sum_{kx_i=0}^{Kx_i-1} \sum_{ky_i=0}^{Ky_i-1} K_i[kx_i, ky_i, c_i, c]\times d_i[(x+kx_i), (y+ky_i), c_i] ,
\end{equation}
where $K_i$ is the kernel, $Kx_i$ is the number of rows in the kernel, and $Ky_i$ is the number of columns in the kernel of the convolution layer i.
To implement \eqref{eq:NN} on hardware, $x_{i+1} \times y_{i+1} \times C_{i+1} \times x_i \times y_i \times C_i$ number of multiplications and $x_{i+1} \times y_{i+1} \times C_{i+1} \times C_i \times Kx_i \times (Ky_i-1)$ number of additions \JS{need} to be performed. In addition to convolution and FC layers, pooling and non-linear activation layers such as  rectified linear unit (ReLU) are present in 
\JS{the DNN} algorithms.

%
\begin{figure}[t]
	\centering
	\includegraphics[width=0.7\textwidth]{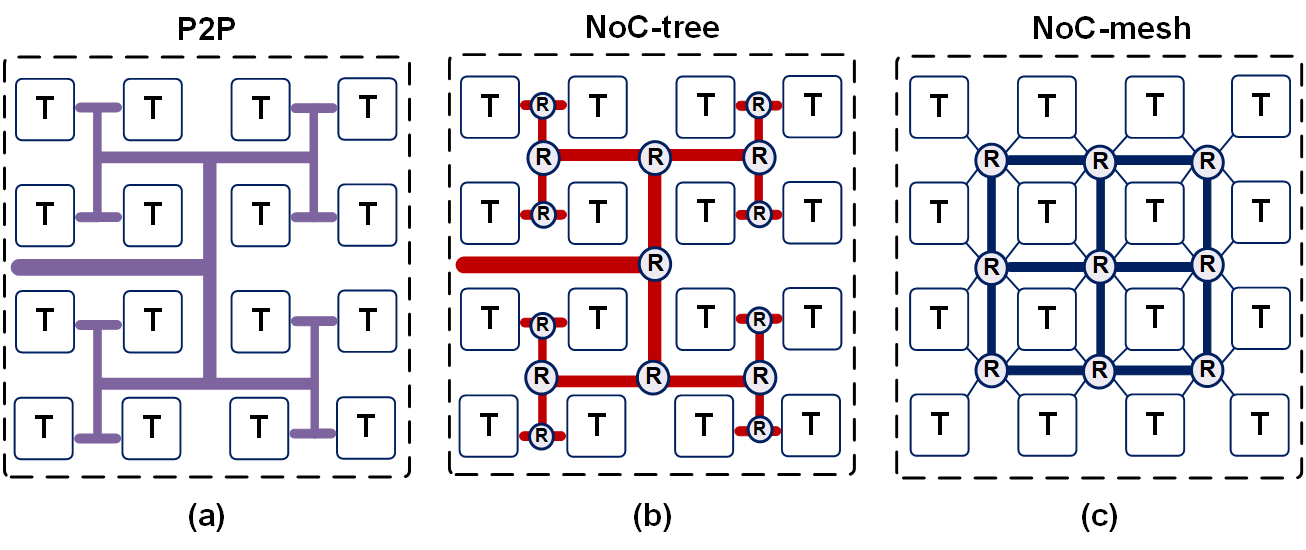}
	\caption{Multi-tiled IMC architecture with routing architectures based on (a) P2P network, (b) NoC-tree, (c) NoC-mesh. NoC-tree is a P2P network with routers at junctions.} 
	\label{fig:inmem}
	\vspace{-5mm}
\end{figure}
\subsection{In-Memory Computing with Crossbars}
DNNs with a large number of weights requires a considerable amount of computations. Conventional architectures separate access of data from the memory and computation in the computing unit. This results in increased computation and data movement, reducing both the throughput and energy-efficiency for DNN inference. In contrast, in-memory computing (IMC) seamlessly integrates computation and memory access in a single unit such as the crossbar~\cite{shafiee2016isaac, song2017pipelayer, krishnan2020interconnect}. Through this, IMC achieves higher energy efficiency and throughput as compared to conventional von-Neumann architectures. 

The IMC technique localizes computation and data memory in a more compact design and enhances parallelism with multiple-row access, resulting in improved performance~\cite{khwa201865nm, C3SRAM}. The data accumulation is achieved through either current or charge accumulation. The size of the IMC subarray usually varies from 64$\times$64 to 512$\times$512.
Along with the computing unit, peripheral circuits such as sample and hold circuit, analog-to-digital converter (ADC), and shift-and-add circuits are used to obtain each DNN layer's result. In this work, we focus on IMC designs based on both SRAM~\cite{khwa201865nm, C3SRAM, yin2019vesti} and ReRAM~\cite{mao2019max2, song2017pipelayer, qiao2018atomlayer, krishnan2020interconnect} crossbars.

\subsection{Interconnect Network}
\begin{figure}[t]
	\centering
    \includegraphics[width=0.7\textwidth]{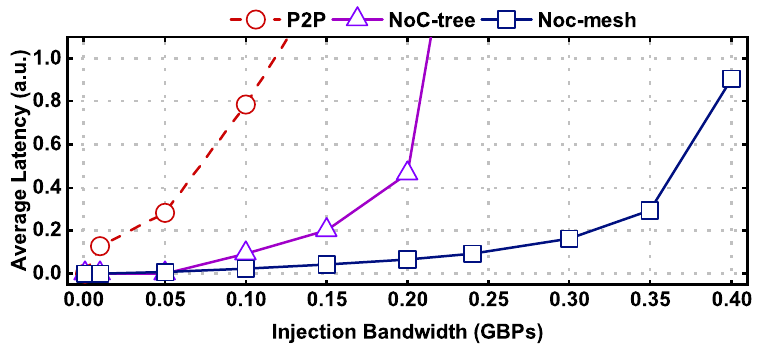}
	\caption{Comparison of average latency among P2P, NoC-tree, and NoC-mesh interconnect for different injection bandwidth~\cite{jiang2013detailed}. NoC topologies show better scalability than P2P interconnect.}
	\label{fig:mesh_vs_htree_latency}
\end{figure}
As discussed in Section~\ref{sec:intro}, the on-chip interconnect is critical to the accelerator performance for DNN acceleration.
There are multiple topologies for Network-on-Chip (NoC). The well-known topologies are mesh, tree, torus, hypercube, and concentrated mesh (c-mesh). NoC with torus topology shows better performance than mesh due to long links between the nodes located at the edges. However, the power consumption by torus is significantly higher than mesh, as shown in~\cite{mirza2007empirical}. Hypercube and c-mesh have a similar disadvantage as a torus. Therefore, only NoC-tree and NoC-mesh are considered in this work. Also, they are the industrial standard for SoCs used in heavy workloads~\cite{jeffers2016intel}.

Figure~\ref{fig:inmem} illustrates representative interconnect schemes of P2P, NoC-tree, and NoC-mesh for multi-tiled IMC architectures.
Each tile consists of several crossbar sub-arrays which perform the IMC operation. Existing implementations of DNN accelerators use both P2P-based~\cite{venkataramani2017scaledeep, kwon2018maeri} and NoC-based~\cite{shafiee2016isaac, krishnan2020interconnect, zhu2020mnsim} interconnect for on-chip communication.
To better understand the performance of different interconnect architectures, we plot the average interconnect latency for a P2P network with 64 nodes, NoC-tree with 64 nodes, and an 8$\times$8 NoC-mesh with X--Y routing as shown in Figure~\ref{fig:mesh_vs_htree_latency}. 
\rev{The NoC utilizes one virtual channel, a buffer size (all input and output buffers) of eight, and three router pipeline stages.}
We observe that for lower injection rates, the performance is comparable for all topologies, while for higher injection rates, NoC performs better in terms of latency. Hence, NoC provides better scalability and performance compared to P2P interconnects.
Moreover, with increasing connection density, injection bandwidth between layers increase due to increased on-chip data movement. Therefore, P2P interconnect performs poorly for DNNs with high connection density. Hence, there is a need for systematic guidance for choosing the optimal interconnect for in-memory acceleration of DNNs.
\rev{Other works such as~\cite{chen2019eyeriss} utilizes three separate NoC for weights, activations and partial sums. Such a design choice results in increased area and energy cost for the interconnect fabric. Furthermore, the three NoCs are under-utilized, resulting in a sub-optimal design choice for acceleration of DNNs.}
\subsection{Analytical Modeling of NoCs}
\rev{Till date, multiple NoC performance analysis techniques have been proposed for SoCs~\cite{ogras2010analytical, mandal2019analytical, mandal2020analytical, kiasari2012analytical, qian2015support}.}
The analytical performance model for an NoC router assumes that the probability distribution of input traffic is in a continuous time domain~\cite{ogras2010analytical}.
However, all transactions in an IMC architecture happen in a discrete clock cycle.
An analytical performance modeling technique for NoCs in the discrete-time domain is proposed in~\cite{mandal2019analytical}.
In this work, we estimate end-to-end communication latency for different DNNs as a function of connection density and the number of neurons of the DNN.
Specifically, we utilize the analytical model for NoC router presented in~\cite{ogras2010analytical} with the modifications for discrete time input~\cite{mandal2019analytical} and extend the model to obtain end-to-end communication latency for NoC-tree and NoC-mesh.

%% file: simulation.tex
\section{Simulation Framework}\label{sec:sim}

There exist multiple simulators that evaluate the performance of DNNs on different hardware platforms~\cite{dong2012nvsim, chen2018neurosim}.
These simulators consider different technologies, platforms, and peripheral circuit modeling while providing less consideration to interconnect.
With the advent of dense DNN structures~\cite{xie2019exploring}, the importance of interconnect cost is higher, as discussed in Section~\ref{sec:intro}. 
In this work, we develop an in-house simulator, where a circuit-level performance estimator of the computing fabric is combined with a cycle-accurate simulator for the interconnect.
The simulator also aims at being versatile by supporting multiple DNN algorithms across different datasets, and various interconnect schemes.

\begin{figure}[t]
	\centering
	\vspace{-4mm}
	\includegraphics[width=0.7\columnwidth]{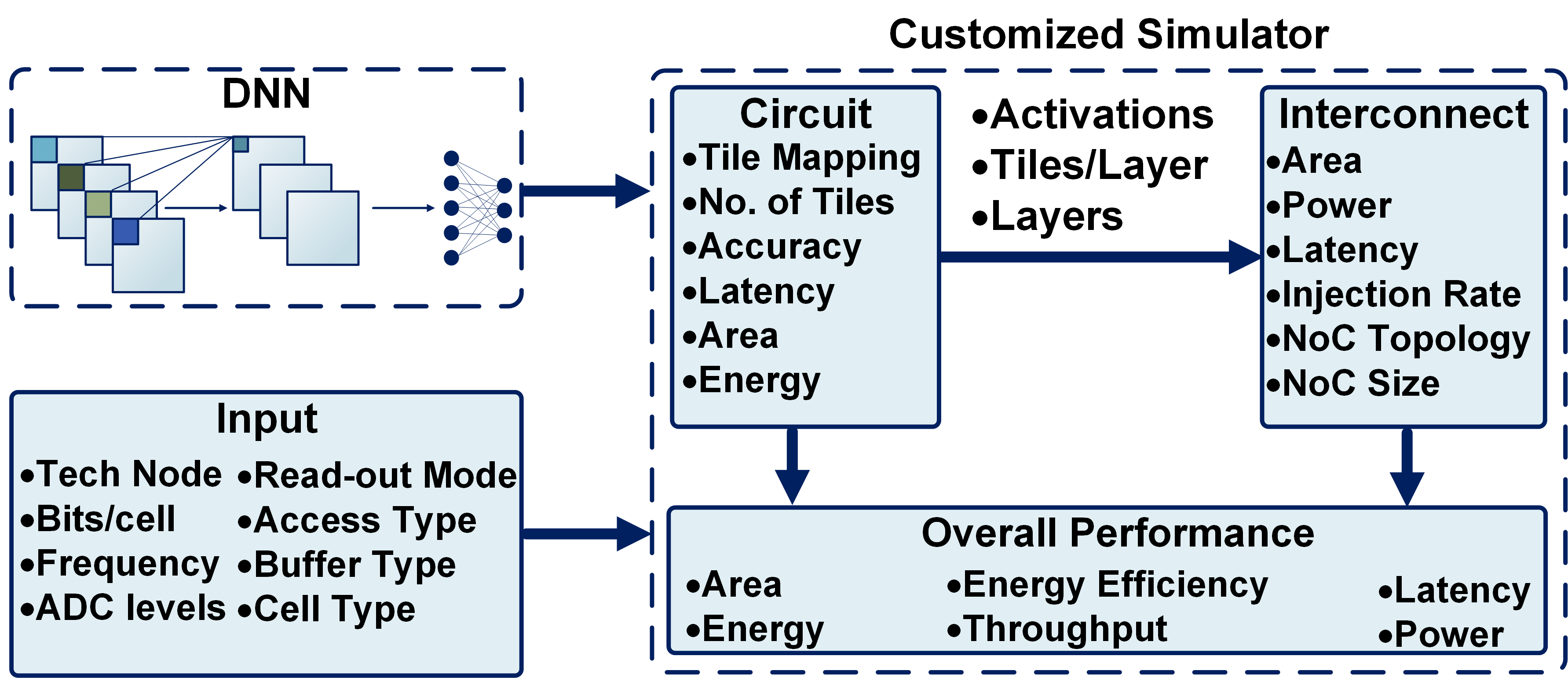}
	\vspace{-2mm}
	\caption{Block-level representation of the proposed architecture simulator.} 
	\label{fig:framework}
	\vspace{-3mm}
\end{figure}

Figure~\ref{fig:framework} shows a  block-level representation of the simulator.
The inputs of the simulator primarily include the DNN structure, technology node, and frequency of operation.
In the proposed simulation framework, any circuit-level performance estimator~\cite{dong2012nvsim, chen2018neurosim} and any interconnect simulator~\cite{jiang2013detailed, garnet} can be plugged in to extract performance metrics such as area, energy, and latency, \JS{proving} a common platform for system-level evaluation.
In this work, we use customized versions of NeuroSim~\cite{chen2018neurosim} for circuit simulation and BookSim~\cite{jiang2013detailed} for cycle-accurate NoC simulation.
\subsection{Circuit-level Simulator: Customized NeuroSim}
The inputs to NeuroSim include the DNN structure indicating the layer size and layer count along with technology node, the number of bits per in-memory compute cell, frequency of operation, read-out mode, etc.
The simulator performs the mapping of the entire DNN to a multi-tiled cross-bar architecture by estimating the number of cross-bar arrays and the number of tiles per layer.
Based on the size of the cross-bar $PE_x$ and $PE_y$, the number of cross-bar arrays is determined by\JS{~\eqref{eqn:PE}}.
\begin{equation}\label{eqn:PE}
\textrm{No. crossbars}= \sum_{i=1}^{N_L} \Big\lceil\frac{(Kx_i \times Ky_i \times C_i)}{(PE_x)_i}\Big\rceil \times \Big\lceil\frac{(C_{i+1}) \times N_{bits}}{(PE_y)_i}\Big\rceil ,
\end{equation}
where $N_{bits}$ is the precision of the weights.
The total number of tiles is calculated as the ratio of the total number of crossbar arrays to the number of crossbar arrays per tile.
Furthermore, the peripheral circuits are laid out, and the complete tile architecture is determined. The peripheral circuits include an ADC, sample and hold circuit, shift and add circuit, and a multiplexer circuit.
However, NeuroSim 
\JS{lacks an accurate estimation of the} interconnect cost in latency, energy, and area. 
Therefore, we replace the interconnect \JS{part} of NeuroSim with customized BookSim.
\JS{We also} extract the performance metrics for tile-to-tile interconnect in NeuroSim and replace it with the BookSim tile-to-tile interconnect. 
With this customization, our circuit simulator only reports performance metrics, such as area, energy, and latency of the computing logic.
It provides the number of tiles per layer, activations, and the number of layers to the interconnect simulator.


\subsection{Interconnect Simulator: Customized BookSim}\label{sec:booksim}

\begin{algorithm}[b]
\caption{Evaluation of interconnect latency through simulation} \label{alg:latency_algorithm}
\SetAlgoLined
\textbf{Input:} Number of layers ($N_L$), Number of tiles in each layer ($T_i$), FPS ($F$), Number of activation in each layer ($A_i$), interconnected topology \\
\textbf{Output:} End-to-end interconnect latency ($L_{routing}$) \\
\For {each layer $i$}  {
\For {each tile $j$ in layer $i-1$} {
\For {each tile $k$ in layer $i$} {
\If {$i > 0$} {
    Compute $\lambda_{i,j,k}$ following Equation~\ref{eq:inj_rate}.
}
}
}
Simulate with interconnect topology and $\lambda_{i,j,k}$\\
Obtain $(l_i)_{sim}$ from the simulator. \\
Calculate $l_i$ following Equation~\ref{eq:end_to_end_layer}.
}
Calculate $L_{comm}^{sim}$ : $L_{comm}^{sim} = \sum_{i=1}^{N_L} l_i$.
\end{algorithm}

\begin{figure}[t]
	\centering
	\includegraphics[width=0.25\columnwidth]{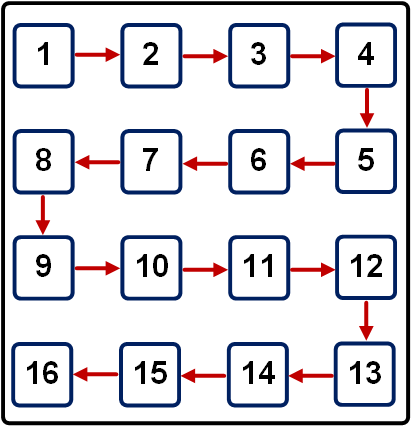}
	\caption{\rev{Tile numbering and placement while mapping the DNN to the IMC architecture. The red arrows show the flow of the data across the tiles.}} 
	\label{fig:dataflow}
\end{figure}

DNNs have varying structures resulting in different traffic loads and data-patterns between the IMC PEs.
To accurately capture the NoC traffic of a given DNN configuration, we 
customize BookSim to evaluate the area, energy, and latency for interconnect, as shown in Figure~\ref{fig:framework}.
\minrev{In the customized version of the BookSim, we enable simulation with non-uniform injection rate.}
We compute the injection rates for each source-destination pair in the multi-tiled architecture.
\rev{The placement of tiles and routers in the IMC architecture has a direct impact on the interconnect performance.
In this work, we incorporate the impact of mapping into the injection matrix calculation.
The mapping of the DNN is performed such that each tile can have at least one layer while no layer is divided between two tiles.
Figure~\ref{fig:dataflow} shows a sixteen tile IMC architecture with the tiles numbered. The red arrows show the data flow in the IMC architecture.
Next, while evaluating the interconnect latency, we create an injection matrix that incorporates the position of the tile into the calculation by calculating the number of hops for each source-destination pair. 
Hence, the injection matrix incorporates the tile placement into the NoC latency calculation. 
Overall, the proposed approach can be generalized to any tile placement.
}
Algorithm~\ref{alg:latency_algorithm} describes the steps performed to compute injection rates and obtain the interconnect latency.
Without loss of generality, we assume that the number of nodes required in the interconnect is equal to the total number of tiles across all layers.
%
%

The injection rate calculation is shown in lines 5--11 of Algorithm~\ref{alg:latency_algorithm}.
\JS{The injection rate is expressed in~\eqref{eq:inj_rate}} from each source to each destination in each layer.
\begin{equation} \label{eq:inj_rate}
    \lambda_{i,j,k} = \frac{A_i \times {N_{bits}} \times FPS}{T_i \times T_{i-1} \times W\times freq}
\end{equation}
where $N_{bits}$, $W$, and $FPS$ represent data precision \JS{and} bus width, and \JS{frames-per-second throughput}, respectively.
In the numerator of \eqref{eq:inj_rate}, we multiply the number of input activations ($A_{i}$) for $i^\mathrm{th}$ layer by ${N_{bits}}$ to obtain the total number of bits to be transferred from $(i-1)^\mathrm{th}$ layer to $i^\mathrm{th}$ layer for one frame of an image.
We further multiply this term with FPS to obtain the total number of bits transferred between layers per second.
Then, we divide this term by the operating frequency ($freq$) to obtain the total number of bits transferred between layers per cycle.
We assume an equal injection rate between all tiles in two consecutive layers.
Therefore, to get the number of bits transferred from one tile to another in two consecutive layers, the denominator in \eqref{eq:inj_rate} includes a multiplication between $T_i$ and $T_{i-1}$.
Thus, we divide the expression obtained so far by $W$ to obtain the injection rate ($\lambda_{i,j,k}$).
The injection rate from every source to every destination is the input to the interconnect simulator.
The interconnect simulator then provides average latency to \JS{complete} all transactions from $(i-1)^\mathrm{th}$ layer to $i^\mathrm{th}$ layer ($(l_i)_{sim}$ cycles).
Next, we multiply this latency with the number of bits from one tile to the next tile to get the total number of cycles required to transfer all data between two consecutive layers.
\JS{Then}, the latency from one layer to the next layer ($l_i$) is given by:
\begin{equation} \label{eq:end_to_end_layer}
    l_i = \frac{(l_i)_{sim}  \times A_i \times {N_{bits}} \times FPS}{freq}
\end{equation}
Finally, we \JS{accumulate} the latency \JS{of all layers} to compute the end-to-end interconnect latency as
\begin{equation}
    L_{comm}^{sim} = \sum_{i=1}^{N_L} l_i
\end{equation}


%% file: noc_performance.tex
\section{Analytical Performance Models for NoCs in IMC Architecture} \label{sec:ana_perf}

In this section, we discuss an analytical approach to estimate NoC performance for IMC architecture.
The analytical performance model of NoCs is primarily useful to overcome longer simulation time incurred by cycle-accurate NoC simulators.
Specifically, we utilize analytical performance models for NoCs to compare the performance of NoC-tree and NoC-mesh for a given DNN.
The analytical model of an NoC router is adopted from the work proposed in~\cite{ogras2010analytical}.
We extend this router model for NoC-tree and NoC-mesh to obtain end-to-end communication latency for different DNNs.
Algorithm~\ref{algo:ana_model} describes the technique to evaluate the communication latency through analytical models.
There are two major steps involved in analyzing the performance of an NoC: 1) Computing injection rate and 2) Computing contention probability matrix.

\noindent\textbf{Computing injection rate matrix ($\Lambda$):} 
First, the injection rate from each source to each destination ($\lambda_{sd}$) for each layer of the DNN is computed through~\eqref{eq:inj_rate}.
\rev{We note that the injection rate calculation incorporates the tile placement as detailed in Section~\ref{sec:booksim}.}
Each NoC router has five ports: North ($N$), South ($S$), East ($E$), West ($W$), and Self ($Se$).
The injection rate at each port $p$ of every router $r$ ($\lambda_p^r, p \in \{N,S,E,W,Se\}$) is computed as:
\begin{equation} \label{eq:inj_rate_router}
    \lambda_p^r = \frac{A_p^r \times N_{bits} \times FPS}{T_l \times T_{l+1} \times W\times freq}
\end{equation}
where $T_l$ denotes the number of tiles in the $l^{\mathrm{th}}$ layer.
$\lambda_p^r$ is a function of the number of activations through each port $p$ of router $r$ ($A_p^r$).
From $\lambda_p^r$, the injection rate matrix for router $r$ ($\Lambda^r$) is computed (as shown in line 5--7 of Algorithm~\ref{algo:ana_model}), where $\Lambda^r = \{ \lambda_{ij}^r \}, 1 \leq i \leq 5,  1 \leq j \leq 5, \lambda_{ij}^r = 0~ \forall i \neq j$.
\begin{algorithm}[t]
\caption{End-to-end latency computation through analytical models} \label{algo:ana_model}
\SetAlgoLined
\SetNoFillComment
\textbf{Input:} Input activation, Number of routers in each layer $l$ ($R_l$), Number of layers ($N_L$) \\
\textbf{Output:} End-to-end communication latency ($L_{comm}$) \\

\For {l = \normalfont{1: $N_L$-1}} {

\For {r = \normalfont{1: $R_l$}} {

\tcc{\textbf{Computing injection rate matrix}}

Compute $A_p^r$ \\
Compute $\lambda_p^r$ using~\eqref{eq:inj_rate_router}  \\
Construct $\Lambda^{r}$ \\

\tcc{\textbf{Computing contention matrix}}

Compute forwarding probability matrix ($F^r$) \\
Compute contention matrix ($C^r$) \\

\tcc{\textbf{Computing average waiting time}}

Compute average queue length ($N^r$) using~\eqref{eq:q_length} \\
Compute average waiting time ($W_{avg}^r$) using~\eqref{eq:avg_w}

}

Compute average latency for the layer ($L_{avg}^l$) using~\eqref{eq:end_latency}
}

$L_{comm}^{ana} = \sum_{l=1}^{N_L} L_{avg}^l$

\end{algorithm}
%

\noindent\textbf{Computing contention matrix ($C$):}
Each element of the contention matrix $C$ ($c_{ij}$) denotes the contention between port $i$ and port $j$.
To compute the contention matrix of router $r$ ($C^r = \{ c_{ij}^r \}$), we first compute forwarding probability matrix $F^r = \{f_{ij}^r\}$.
$f_{ij}^r$ denotes the probability 
of a packet that arrived at the port $i$ of the router $r$ to be forwarded to the port $j$, 
and is computed as shown in~\eqref{eq:fwd_prob}~\cite{ogras2010analytical}.
\begin{equation} \label{eq:fwd_prob}
    f_{ij}^r = \frac{\lambda_{ij}^r}{\sum_{k=1}^{5} \lambda_{jk}^r}
\end{equation}
The contention probability between port $i$ and port $j$ of the router $r$ is computed as $c_{ij}^r = \sum_{k=1}^{5} f_{ik}^r f_{jk}^r$.
Line 10-11 of Algorithm~\ref{algo:ana_model} shows the computation of the contention matrix.

Next, the average queue length of each port of the router $r$ ($N^r$) is computed through the technique described in~\cite{ogras2010analytical}.
\begin{equation} \label{eq:q_length}
    N^r = (I - t\Lambda^r C^r)^{-1} \Lambda^r R ,
\end{equation}
where $t$ is the service time of the router, and we assume $t=1$ for our evaluation.
$R$ is the average residual time and is calculated assuming that the packets arrive in discrete clock cycles~\cite{mandal2019analytical}.
Waiting time of the packets at each port of the router $r$ is computed as $W^r = N^r (\Lambda^r)^{-1}$.
End-to-end average latency for each layer $l$ ($L_{avg}^l$) is obtained by averaging the waiting time through all 5 ports ($W_{avg}^r$) of router $r$ and then adding across all routers, as shown in \eqref{eq:avg_w} and \eqref{eq:end_latency}.
\begin{align} \label{eq:avg_w}
    W_{avg}^r = \frac{1}{5} \sum_{p=1}^{5} W^r_p \\ \label{eq:end_latency}
    L_{avg}^l = \sum_{r=1}^{R} W_{avg}^r
\end{align}
Finally, total communication latency ($L_{comm}^{ana}$) is obtained by adding end-to-end average latency for each layer $l$ as:
\begin{equation} \label{eq:final_model}
    L_{comm}^{ana} = \sum_{l=1}^{N_L} L_{avg}^l
\end{equation}

%% file: Interconnect.tex
\section{Connection-centric Architecture}\label{sec:arch}
\begin{figure}[t]
	\centering
	\includegraphics[width=0.6\textwidth]{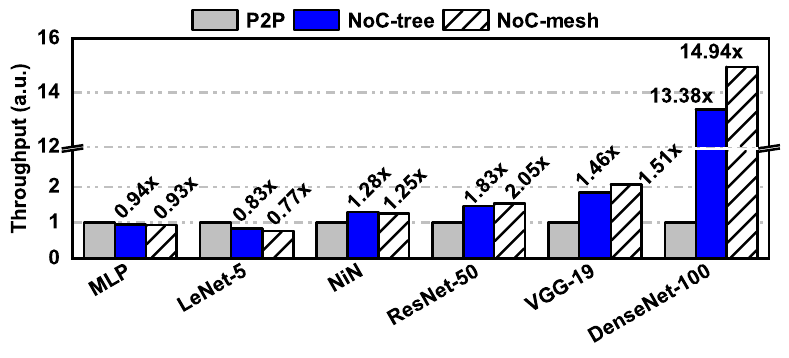}	
	\caption{Throughput comparison for three interconnect topologies (P2P, NoC-tree, and NoC-mesh) for SRAM-based IMC architecture, normalized to P2P, for different DNNs. NoC shows superior performance and scalability than P2P-based network.}
	\label{fig:fps_comp}
\end{figure}
In this section, we first discuss a multi-tiled SRAM-based IMC architecture with three different interconnect topologies, namely, P2P, NoC-tree, and NoC-mesh at the tile level.
We perform a comprehensive analysis of these three interconnect-based SRAM IMC architectures for different DNNs using the simulation framework described in Section~\ref{sec:sim}.
Based on the analysis, we show the need for an NoC-based heterogeneous interconnect IMC architecture for efficient DNN acceleration.
We assume all weights are stored on-chip to avoid any DRAM access. 
\rev{The weights are loaded pre-execution and stored on-chip. 
The inputs are then loaded, and the computation is performed. 
There is no re-loading of intermediate results or weights from the off-chip memory during the execution of the DNN. 
The SRAM buffer is designed large enough to hold the intermediate results on-chip rather than moving them off-chip. 
Multiple inferences of the images can be performed using one pre-execution loading of the weights. Hence, we do not consider the initial loading of the weights into the energy calculation, consistent with prior work~\cite{shafiee2016isaac, song2017pipelayer} compared in the manuscript.}
In addition, we adhere to layer-by-layer design instead of a layer-pipelined design, since a pipelined design introduces pipeline bubbles in the execution flow and complicates the control logic~\cite{qiao2018atomlayer}.
%
\subsection{Design Space Exploration}\label{sec:dsn}
We evaluate different performance metrics for a wide range of DNNs with P2P, NoC-tree, and NoC-mesh-based interconnect for SRAM-based IMC architectures.
\rev{We consider routers with five ports, one virtual channel for NoCs and X--Y routing for NoC-mesh for this evaluation.}
To facilitate fair comparison, we normalize the throughput of the hardware architectures with three interconnect topologies to that of P2P interconnect.

Figure~\ref{fig:fps_comp} shows the throughput comparison for different DNNs. For low connection density DNNs such as MLP and LeNet-5~\cite{lecun1998gradient}, the choice of interconnect does not make a significant difference to the throughput, due to low data movement between different tiles of the IMC architecture. However, P2P interconnect results in 1.25$\times$ and 2$\times$ higher area cost than NoC-tree for MLP and LeNet-5, respectively.
Hence, NoC-tree provides better overall performance than P2P for both MLP and LeNet-5.
We further analyze dense DNNs such as NiN~\cite{lin2013network}, VGG-19, ResNet-50~\cite{he2016deep} and DenseNet-100~\cite{huang2017densely}. 
The performance comparison shows that the NoC-tree and NoC-mesh-based IMC architectures perform better than the P2P-based architectures (up to 15$\times$ for DenseNet-100).
Since higher connection density of the DNNs results in increased on-chip data movement, the routing latency dominates the end-to-end latency. We see a similar trend with ReRAM-based IMC architectures with similar throughput for MLP and 15$\times$ improvement in throughput for DenseNet-100.
Through this, we establish that the performance of the P2P-based IMC architecture (SRAM- or ReRAM-based) diminishes with increasing connection density. In contrast, the performance of the NoC-based (tree, mesh) IMC architecture scales better (Figure~\ref{fig:fps_comp}).
\begin{figure}[t]
	\centering
	\includegraphics[width=0.6\textwidth]{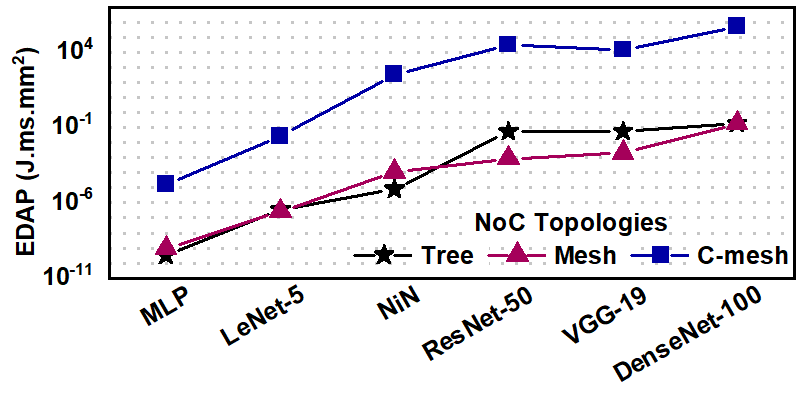}
	\vspace{-2mm}
    \caption{\rev{Comparison of energy-delay-area product (EDAP) of NoC-tree, NoC-mesh, and c-mesh for different DNNs.}}
	\label{fig:noc_edap}
\end{figure}

\noindent\textbf{\rev{Exploration of other NoC topologies:}}
\rev{Apart from tree and mesh, the other commonly known NoC topologies include c-mesh, hypercube, and torus.
These topologies utilize more resources in terms of routers and links to reduce communication latency.
However, the usage of more resources increases power consumption and the area of the NoC.
For example, we performed experiments with c-mesh topology for different DNNs.
Figure~\ref{fig:noc_edap} compares energy-delay-area product (EDAP) of mesh-, tree- and c-mesh-based NoC for different NoC.
We observe that while mesh- and tree-NoC provides comparable EDAP, the same for c-mesh is a minimum of five orders of magnitude higher than mesh- and tree-NoC.}

\subsection{Hardware Architecture}\label{sec:hardware_arch}
\begin{figure}[t]
	\centering
	\includegraphics[width=0.6\textwidth]{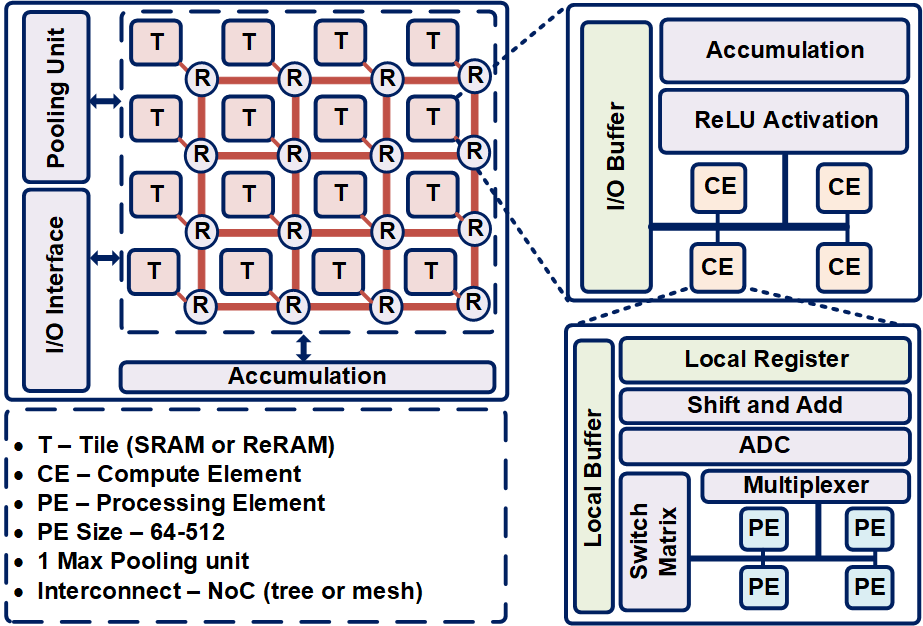}
	\caption{NoC-based heterogeneous interconnect IMC architecture. A three-level interconnect scheme consisting of NoC (tree or mesh) between tiles, P2P network between CEs, and bus between PEs.}
	\label{fig:htree_arch}
\end{figure}
Based on the conclusions from Section~\ref{sec:dsn}, we derive an NoC-based heterogeneous interconnect IMC architecture for DNN acceleration.
Figure~\ref{fig:htree_arch} shows the hardware architecture which employs the heterogeneous interconnect system.

The proposed architecture is divided into a number of tiles, with each tile having a set of computing elements (CE). The tile architecture includes non-linear activation units, I/O buffer, and accumulators to manage data transfer efficiently.  
Each CE further consists of multiple processing elements (PE) or crossbar arrays, multiplexers, buffers, a sense amplifier, and flash ADCs. The ADC precision is set to four bits such that there is minimum or no accuracy degradation for DNNs. In addition, the architecture does not utilize a digital-to-analog (DAC) converter; instead, it uses sequential signaling to represent multi-bit inputs~\cite{peng2019inference}.
The proposed heterogeneous tile architecture can be used for both SRAM and ReRAM (1T1R) technologies. However, the peripheral circuits change based on the technology.
In this work, we choose a homogeneous tile design consisting of four CEs and a CE structure consisting of four PEs. We evaluate both SRAM- and ReRAM-based IMC architectures for PE sizes varying from 64$\times$64 to 512$\times$512. We sample 8 DNNs (LeNet, NiN, SqueezeNet, ResNet-152, ResNet-50, VGG-16, VGG-19, and DenseNet-100) and a crossbar size of 256$\times$256 provides the lowest EDAP for 75\% of the DNNs. Hence, in this work, we choose 256$\times$256 as the crossbar size for both SRAM- and ReRAM-based IMC architectures.
To maximize performance, the architecture uses heterogeneous interconnects. It employs the NoC-based interconnect on the global tile-level with a P2P interconnect (H-Tree) at the CE-level and bus at the PE-level due to significantly lower data volume. For low data volume, the NoC-based interconnect provides marginal performance gain while increasing energy consumption.

%

%% file: expt_results.tex
\section{Experiments and Results}\label{sec:expt}

\subsection{Experimental Setup}

We consider an IMC architecture (Figure~\ref{fig:htree_arch}) with a homogeneous tile structure (SRAM, ReRAM) and one NoC router per tile. Table~\ref{tab:metrics} summarizes the design parameters considered.
We report the end-to-end latency, chip area, and total energy obtained for a PE size of 256$\times$256 for each of the DNNs using the simulation framework discussed in Section~\ref{sec:sim}.
We incorporate conventional mapping~\cite{shafiee2016isaac}, IMC SRAM bitcell/array design from~\cite{khwa201865nm} and 1T1R ReRAM bitcell/array properties from~\cite{chen2018neurosim}. The IMC compute fabric utilizes a parallel read-out method. We utilize the same crossbar array size of 256$\times$256 for both SRAM and ReRAM-based IMC architectures. All rows of the IMC crossbar are asserted together, analog MAC computation is performed along the bitline, and the analog voltage/current is digitized with a 4-bit flash ADC at the column periphery. We perform an extensive evaluation of the IMC architecture with both SRAM-based and ReRAM-based PE arrays for both NoC-tree and NoC-mesh.
\rev{Unless specified, the NoC utilizes one virtual channel, a buffer size (all input and output buffers) of eight, and three router pipeline stages.}
\begin{table}[h]
\caption{Summary of design parameters} \label{tab:metrics}
\renewcommand*{\arraystretch}{0.9}
\centering
\begin{tabular}{@{}ll|ll@{}}
\toprule
\begin{tabular}[c]{@{}l@{}}PE array size\end{tabular}                         & 256$\times$256 & Read-out Method                                              & \begin{tabular}[c]{@{}l@{}}Parallel\end{tabular} \\ \midrule
\begin{tabular}[c]{@{}l@{}}Technology node \end{tabular}                   & 32nm          & Flash ADC resolution                                            & 4 bits                                                       \\ \midrule
Cell levels                                                                       & 1 bit/cell     & \begin{tabular}[c]{@{}l@{}}Operating frequency\end{tabular} & 1 GHz                                                        \\ \midrule
\begin{tabular}[c]{@{}l@{}}Data precision\end{tabular} & 8 bits              & NoC bus width                                                 & 32                                                         \\
\bottomrule
\end{tabular}
\end{table}
%
\subsection{\rev{Evaluation of NoC Analytical Model}}
\begin{figure}[t]
	\centering
	\includegraphics[width=0.6\textwidth]{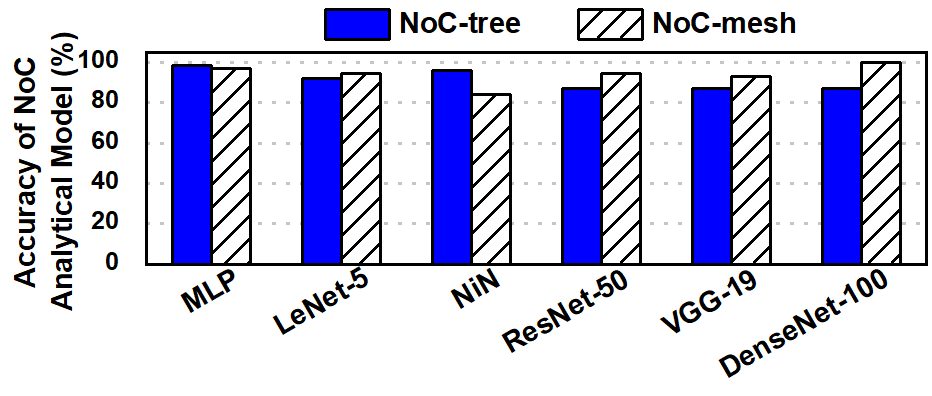}
	\caption{Accuracy of NoC analytical model for NoC-mesh and NoC-tree with respect to cycle-accurate simulator~\cite{jiang2013detailed}.}
	\label{fig:model_verification}
\end{figure}
\begin{figure}[t]
	\centering
	\includegraphics[width=0.6\textwidth]{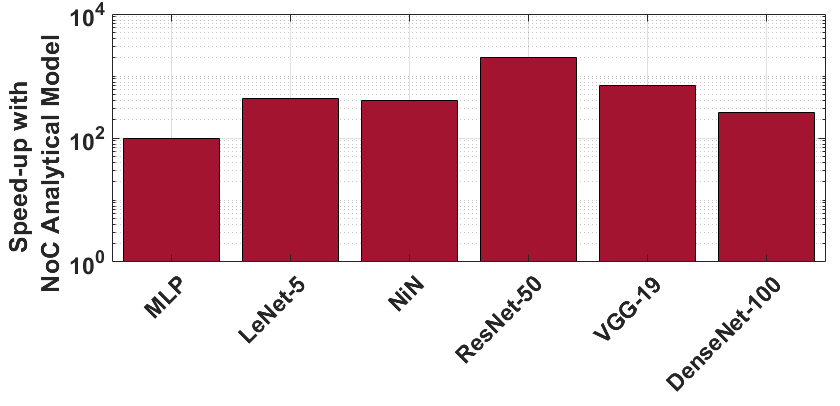}
	\caption{\rev{Speed-up (in NoC simulation) with NoC analytical models with respect to cycle-accurate NoC simulation for different DNNs with mesh-NoC.}}
	\label{fig:noc_speedup}
\end{figure}

Figure~\ref{fig:model_verification} shows the accuracy of the analytical model (presented in Algorithm~\ref{algo:ana_model} in Section~\ref{sec:ana_perf}) to estimate the end-to-end communication latency with both NoC-tree and NoC-mesh.
We observe that the accuracy is always more than 85\% for different DNNs.
On an average, the NoC analytical model achieves 93\% accuracy with respect to cycle-accurate NoC simulation~\cite{jiang2013detailed}.
\rev{Moreover, we achieve 100$\times$-2000$\times$ speed-up with the NoC analytical model with respect to cycle-accurate NoC simulation. 
Figure~\ref{fig:noc_speedup} shows the speed-up for different DNNs with mesh-NoC. 
This speed-up is useful to perform design space exploration by considering various sizes of PE arrays and other NoC topologies.
Due to the high speed-up in NoC performance analysis , we achieve 8$\times$ speed-up in overall performance analysis with respect to the framework which uses cycle-accurate NoC simulation.}

\subsection{\rev{Analysis on Traffic Congestion in NoC}}

\rev{In this section, we present an analysis on traffic congestion in NoC for various DNNs.
To this end we discuss about average queue length of different buffers in the NoC and worst case communication latency.}

\begin{figure}[t]
	\centering
	\includegraphics[width=0.6\textwidth]{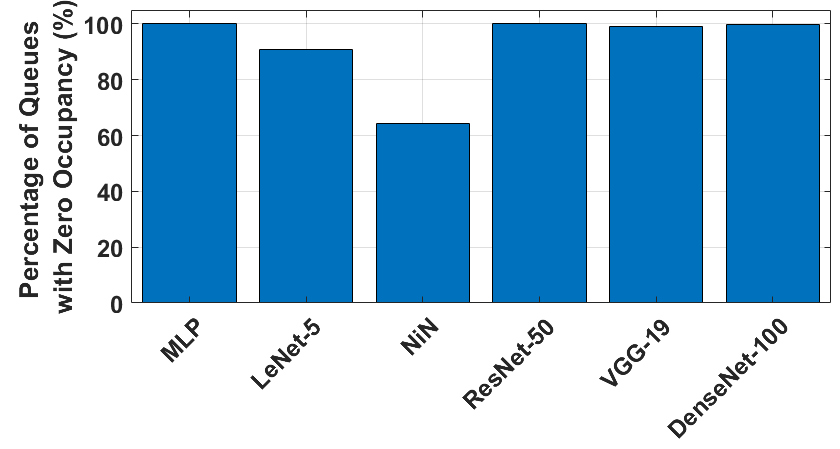}
	\caption{\rev{Percentage of queues with zero occupancy when a new flit arrives.}}
	\label{fig:zero_occupancy}
\end{figure}

\noindent\rev{\textbf{Analysis of the average queue length:}}
\rev{Furthermore, we investigated the average queue length at different ports of different routers in the NoC through a cycle-accurate NoC simulator. We performed this experiment with mesh-NoC considering the configuration parameters shown in Table~\ref{tab:metrics}. 
Figure~\ref{fig:zero_occupancy} shows that 64\%-100\% of the queues contain no flit when a new flit arrives for different DNNs. The percentage of queues with zero occupancy for LeNet-5 and NiN is 91\% and 65\%, respectively. These two DNNs utilize fewer number of routers, which results in less parallelism in data communication. However, we note that determining the optimal number of routers for a given DNN is not a scope of this work.}

\begin{figure}[t]
\centering
\begin{subfigure}{0.5\textwidth}
  \centering
  \includegraphics[width=1\textwidth]{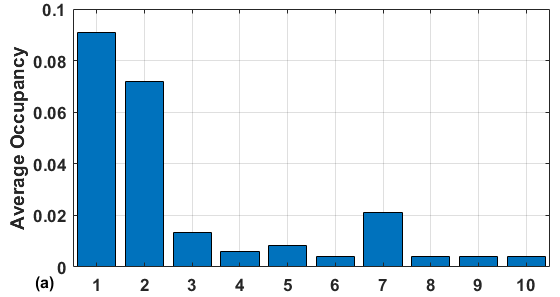}
\end{subfigure}%
\begin{subfigure}{0.5\textwidth}
  \centering
  \includegraphics[width=1\textwidth]{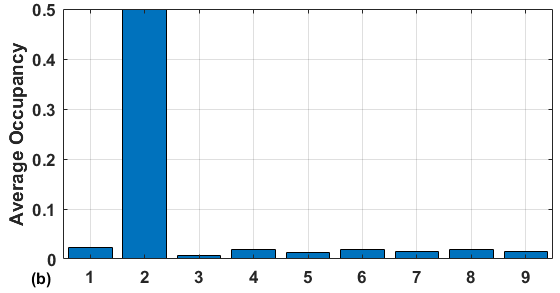}
\end{subfigure}
\caption{\rev{Average Occupancy of Queues with non-zero length for (a) NiN, (b) VGG-19.}}
\label{fig:zero_occup_dnn}
\end{figure}

\rev{Figure~\ref{fig:zero_occup_dnn} shows the average queue length for NiN and VGG-19 for the queues with non-zero length when a new flit arrives to the queues. We observe that the average queue length varies from 0.004-0.5 for these DNNs. Average queue length is very low in these cases since the injection rate to the queues are less, and NoC introduces a high degree of parallelism in data transmission between routers.}

\noindent\rev{\textbf{Analysis of the worst case latency:}}
\rev{Furthermore, we extracted the worst-case latency ($L_{max}$) for different source to destination pairs of different DNNs with mesh-NoC. We compared $L_{max}$ of each source to destination pair with corresponding average latency ($L_{avg}$).
Then we compute mean absolute percentage deviation (MAPD) of $L_{max}$ from $L_{avg}$ as the equation below.}
\begin{equation}
    \rev{MAPD =100 \times \frac{1}{N} \sum_{i=1}^{N} \frac{(L_{max}^i - L_{avg}^i)}{L_{avg}^i}}
\end{equation}
\rev{Where $N$ is the total number of source to destination pairs with non-zero average latency. $L_{max}^i$ and $L_{avg}^i$ are the worst-case latency and the average latency respectively of $i^\mathrm{th}$ source to destination pair.
Table~\ref{tab:worst_case_all} shows the mean absolute percentage deviation for different DNNs.
We observe that the deviation is insignificant, except for LeNet-5 and NiN.
The deviations for these two networks are 9.13\% and 20.76\%, respectively.
}

\begin{figure}[t]
\centering
\begin{subfigure}{0.5\textwidth}
  \centering
  \includegraphics[width=1\textwidth]{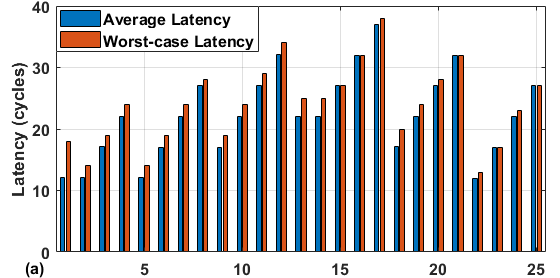}
\end{subfigure}%
\begin{subfigure}{0.5\textwidth}
  \centering
  \includegraphics[width=1\textwidth]{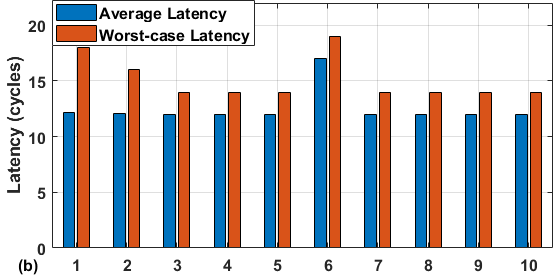}
\end{subfigure}
\caption{\rev{Comparison between average latency and worst-case latency for source to destination pairs with non-zero latency for (a) LeNet-5 and (b) NiN.}}
\label{fig:worst_case}
\end{figure}

\rev{Furthermore, in Figure~\ref{fig:worst_case} we show the absolute difference between the worst-case latency and the average latency for LeNet-5 and NiN for different source to destination pairs with non-zero latency. The maximum difference is 6 cycles both for LeNet-5 and NiN. This analysis shows that the worst-case latency has very less deviation from the average latency. Therefore, the studies of average queue length and worst-case latency confirm that there is no congestion in the NoC.}

\begin{table}[t]
\caption{\rev{Mean absolute percentage deviation ($MAPD$) of worst-case NoC latency from average NoC for different DNNs.}} \label{tab:worst_case_all}
\begin{tabular}{l|l|l|l|l|l|l}
\hline
DNNs     & MLP & LeNet-5 & NiN   & ResNet-50 & VGG-19 & DenseNet-100 \\ \midrule
MAPD(\%) & 0   & 9.13    & 20.76 & 0         & 0.14   & 0            \\ \hline
\end{tabular}
\end{table}

\subsection{Guidance on Optimal Choice of Interconnect}
\subsubsection{Empirical Analysis}
We compare the performance of the IMC architecture using both NoC-tree and NoC-mesh for both SRAM and ReRAM-based technologies. We perform the experiments for representative DNNs. MLP, LeNet-5, and NiN depict low connection density DNNs; ResNet-50, VGG-19, and DenseNet-100 depict high connection density DNNs. We report throughput and the product of energy consumption, end-to-end latency, and area (EDAP) of the IMC architectures. 
\rev{EDAP is used as the metric to guide the optimal choice for the interconnect for IMC architectures.}

\begin{figure}[t]
	\centering
	\includegraphics[width=1\textwidth]{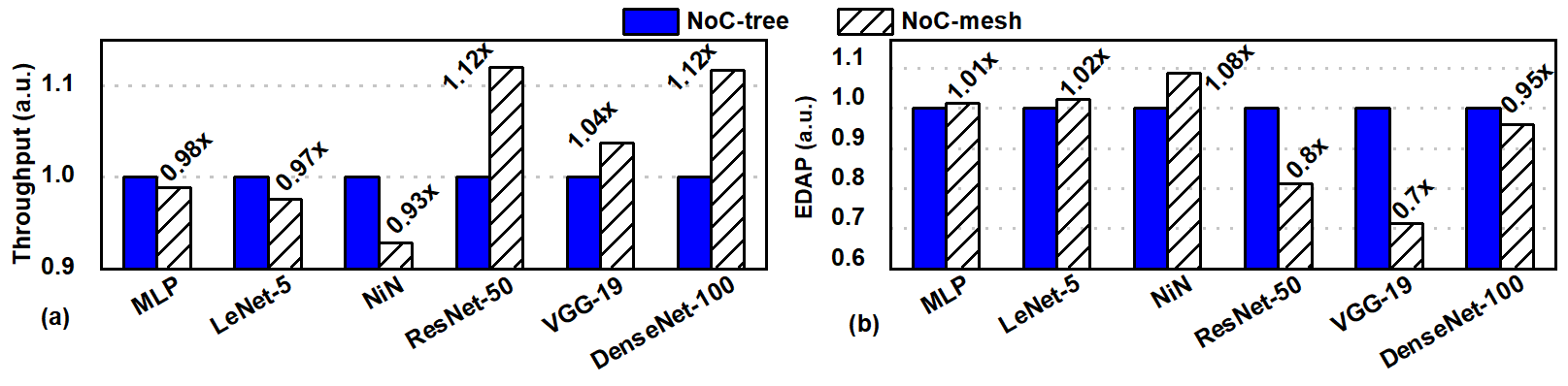}
	\caption{(a) Normalized throughput and (b) normalized EDAP of NoC-tree and NoC-mesh-based on-chip interconnect for \textbf{SRAM-based} IMC architecture for different DNNs. Dense DNNs favor NoC-mesh while NoC-tree is sufficient for shallow DNNs.}
	\label{fig:perf2_SRAM}
\end{figure}

Figure~\ref{fig:perf2_SRAM}(a) shows the ratio of the throughput of the SRAM-based IMC architecture using NoC-tree and NoC-mesh interconnect. We normalize the throughput values with respect to that of NoC-tree.
NoC-tree performs better than the NoC-mesh for DNNs with low connection density. This is because of the reduced injection bandwidth into the interconnect. In addition, while NoC-mesh provides lower interconnect latency than NoC-tree, it comes at an increased area and energy cost. 
However, NoC-mesh performs better for DNNs with high connection density. The improved performance stems from the reduced interconnect latency for high injection rates of data into the interconnect. The reduction in latency is much higher than the additional overhead due to both area and energy of NoC-mesh.
\begin{figure}[t]
	\centering
	\vspace{-2mm}
	\includegraphics[width=1\textwidth]{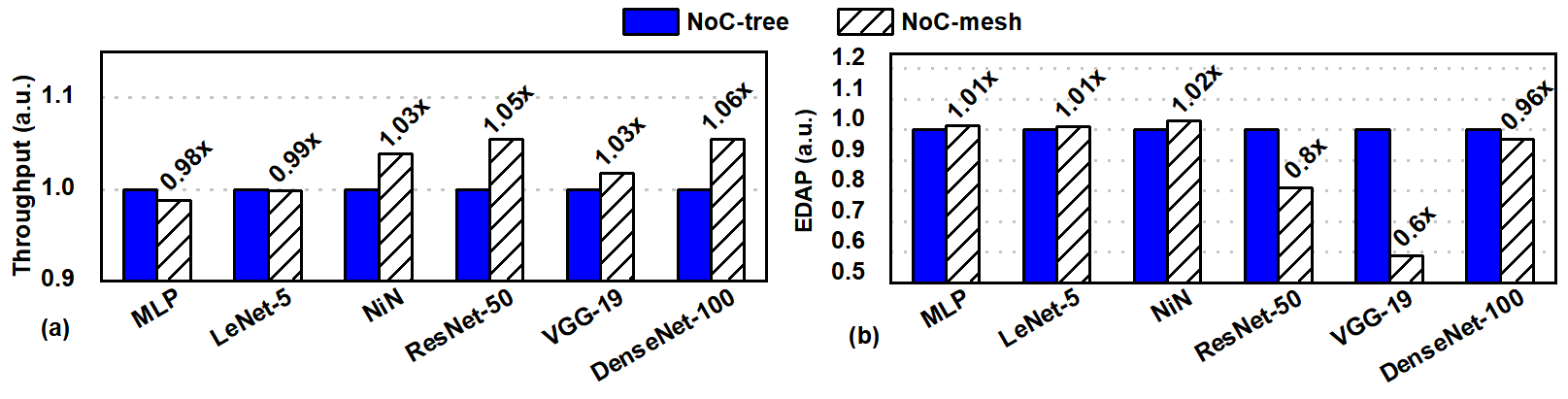}
	\vspace{-4mm}
	\caption{(a) Normalized throughput and (b) normalized EDAP of NoC-tree and NoC-mesh-based on-chip interconnect for \textbf{ReRAM-based} IMC architecture for different DNNs. Dense DNNs favor NoC-mesh while NoC-tree is sufficient for shallow DNNs.}
	\label{fig:perf2_RRAM}
	\vspace{-4mm}
\end{figure}

To better understand the performance, we report the EDAP for the SRAM-based IMC architecture.
Figure~\ref{fig:perf2_SRAM}(b) shows normalized EDAP of the NoC-tree and NoC-mesh for both low and high connection density DNNs.
DNNs with low connection density have significantly lower EDAP for NoC-tree than that with NoC-mesh. Such an improved EDAP performance for NoC-tree complements the observation for throughput. At the same time, for DNNs with high connection density, the EDAP of NoC-mesh is lower than that of the NoC-tree for IMC architectures. A similar observation is seen for ReRAM-based IMC architectures as shown in Figure~\ref{fig:perf2_RRAM}(a) and Figure~\ref{fig:perf2_RRAM}(b). In contrast to the SRAM-based IMC architecture, NiN provides better performance in throughput for the NoC-mesh interconnect. At the same time, NoC-tree provides better EDAP compared to NoC-mesh, similar to that of the SRAM-based IMC architecture.

\begin{figure}[t]
	\centering
	\includegraphics[width=1\textwidth]{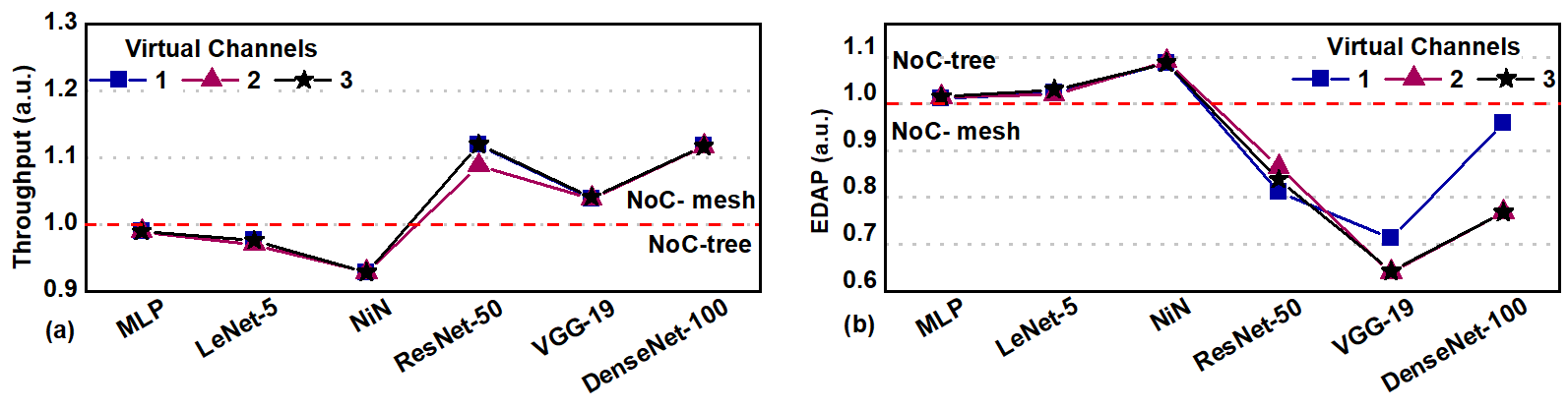}
	\caption{\rev{Assessment of (a) throughput and (b) EDAP between NoC-tree and NoC-mesh with different numbers of virtual channels for different DNNs. The throughput is normalized to that of NoC-tree. The preferred NoC topology for optimal performance is shown for the regions above and below the red line.}}
	\label{fig:vc}
	\centering
	\includegraphics[width=1\textwidth]{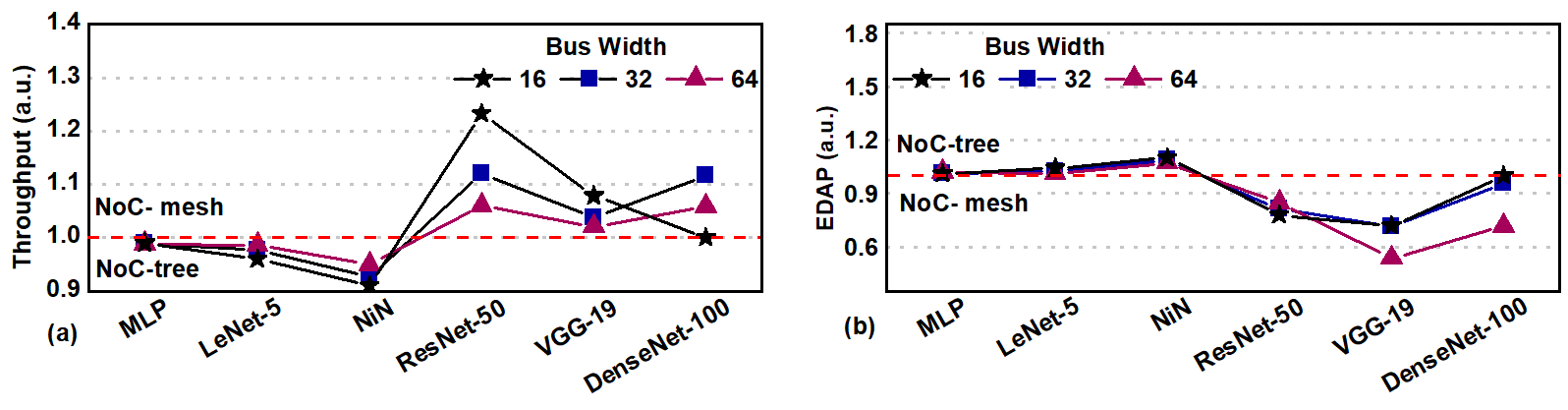}
	\caption{\rev{Assessment of (a) throughput and (b) EDAP between NoC-tree and NoC-mesh with different bus width for different DNNs. The throughput is normalized to that of NoC-tree. The preferred NoC topology for optimal performance is shown for the regions above and below the red line.}}
	\label{fig:bus_width}
\end{figure}

\rev{Furthermore, we performed another two sets of experiments with NoC-mesh and NoC-tree by varying the number of virtual channels and bus-width.
In this case, we consider ReRAM-based IMC architectures.
Figure~\ref{fig:vc} shows the comparison with different numbers of virtual channels, and Figure~\ref{fig:bus_width} shows the comparison with different bus width of the NoC.
We observe similar trends for different DNNs with a different NoC configurations.
}

\rev{Since the injection rate to the input buffer of the NoC is always low (less than one packet in 100 cycles), increasing the number of virtual channels does not alter the inference latency significantly. Therefore, throughput remains similar (for all DNNs) both for NoC-tree and NoC-mesh with an increasing number of virtual channels. However, the area and power of both NoC-mesh and NoC-tree increase proportionally with an increasing number of virtual channels. Therefore, the normalized EDAP (EDAP of mesh-NoC divided by the EDAP of tree-NoC) is similar for all DNNs with different numbers of virtual channels.}

\rev{While we change the bus width of the NoC, the latency increases (decreases) with decreasing (increasing) bus width proportionally, i.e., the latency with a bus width of 32 is twice than the latency with bus width of 64. Moreover, the area and power of the NoC increases (decreases) with increasing (decreasing) bus width proportionally. Therefore, the normalized EDAP is similar for all DNNs with different NoC bus widths. Consequently, for all configurations, we obtain exactly the same guidance on the choice of NoC for different DNNs. Therefore, the guidance is consistent across different parameters of NoCs.}

%
\subsubsection{Theoretical Analysis}
We utilize the analytical model in Section~\ref{sec:ana_perf} and the experimental results described in Figure~\ref{fig:perf2_SRAM} and Figure~\ref{fig:perf2_RRAM} to provide guidance on the optimal choice of interconnect for IMC architectures.
The injection rate at each port of an NoC router for each layer of the DNN is expressed in~\eqref{eq:inj_rate_router}. The numerator of~\eqref{eq:inj_rate_router} denotes the total data volume between $i^\mathrm{th}$ layer and $(i+1)^\mathrm{th}$ layer for each port of the router per cycle. This is divided by $(T_i \times T_{i+1} \times W)$ to obtain the injection rate from for each port of every router as detailed in  Section~\ref{sec:ana_perf}. For a fixed NoC-based IMC architecture, target throughput ($FPS$), frequency of operation ($freq$), and bus width ($W$) remain constant. Hence, from~\eqref{eq:inj_rate_router} we obtain,
\begin{equation} \label{eq:crate}
    \lambda_{i} \propto \frac{A_i \times {N_{bits}}}{T_i \times T_{i+1}}
\end{equation}
Let the connection density for $i^\mathrm{th}$ layer be $\rho_i$ and the number of neurons be $\mu_i$. Data volume between $i^\mathrm{th}$ and $(i+1)^\mathrm{th}$ layer is proportional to the product of $\rho_i$ and $\mu_i$, as shown in~\eqref{eq:d_vol}.
\begin{equation} \label{eq:d_vol}
    (A_i \times {N_{bits}}) \propto (\rho_i \times \mu_i)
\end{equation}
Additionally, the number of tiles in $i^\mathrm{th}$ layer is directly proportional to $\mu_i$.
Hence, from~\eqref{eq:crate} and~\eqref{eq:d_vol} we get,
\begin{align} \label{eq:d_vol_2}
    \lambda_{i} & \propto \frac{\rho_i \times \mu_i}{\mu_i \times \mu_{i-1}} =\frac{\rho_i}{\mu_{i-1}}
\end{align}
Generalising~\eqref{eq:d_vol_2}, we obtain,
\begin{equation} \label{eq:general}
    \lambda \propto \frac{\rho}{\mu}
\end{equation}
\begin{figure}[t]
	\centering
	\includegraphics[width=0.6\textwidth]{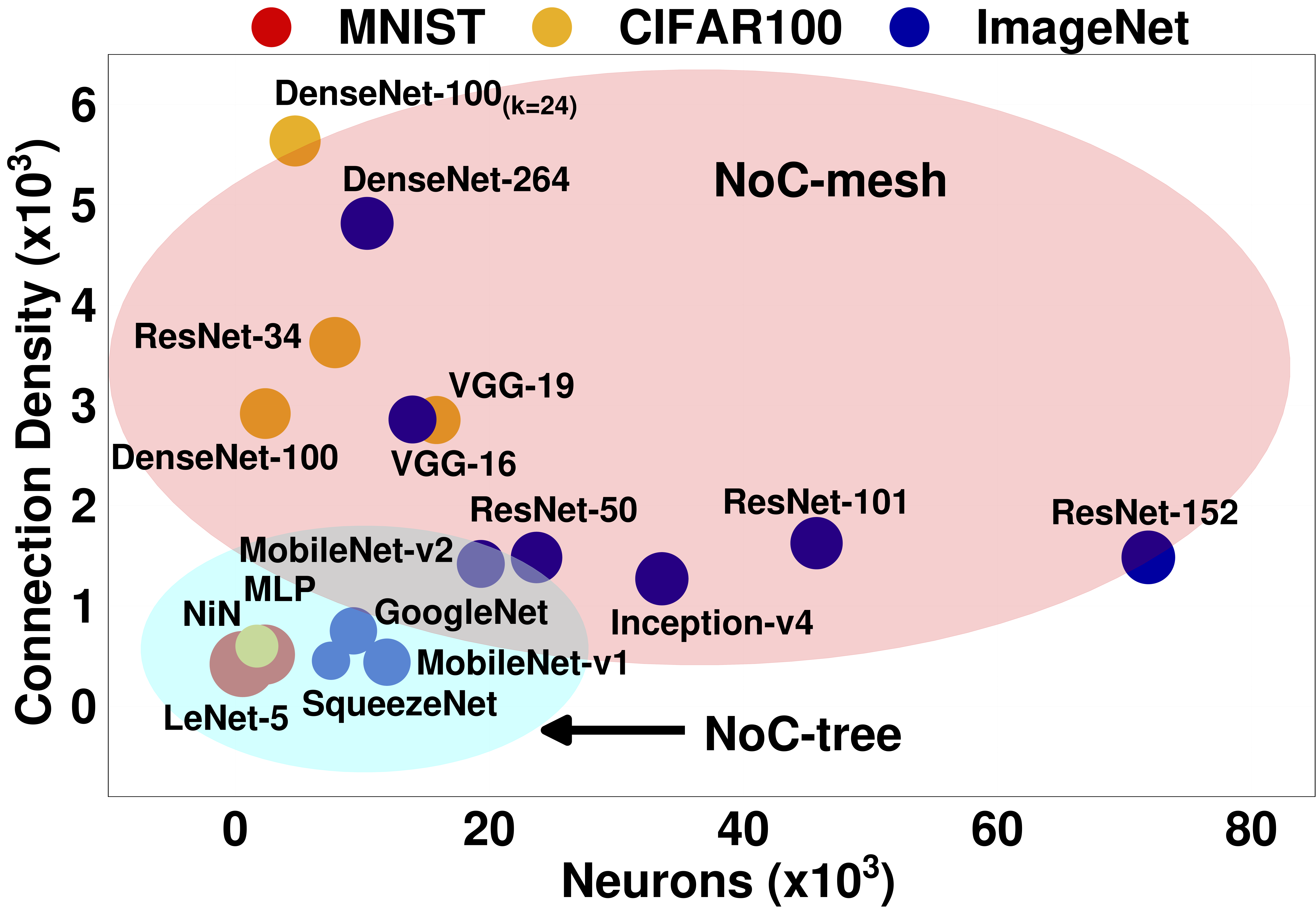}
    \caption{Optimal NoC topology for IMC architectures for different DNNs.} 
	\label{fig:guide}
	\vspace{-3mm}
\end{figure}
Therefore, the injection rate is directly proportional to the connection density and inversely proportional to the number of neurons of the DNN.
Figure~\ref{fig:guide} presents the preferred regions for NoC-tree and NoC-mesh for best throughput for different DNNs with IMC architectures. If the connection density of a DNN is more than 2$\times$$10^3$, then NoC-mesh is suitable. If the connection density is less than 1$\times$$10^3$, then NoC-tree is appropriate.
Both NoC-tree and NoC-mesh are suitable for the DNNs with connection density in the range of 1$\times$$10^3$-2$\times$$10^3$ (the region where red and blue ovals overlap in Figure~\ref{fig:guide}). 


\subsection{Comparison with state-of-the-art architectures}
Table~\ref{tab:final_result} compares the proposed architecture with state-of-the-art DNN accelerators.
Prior works show the efficacy of their ReRAM-based IMC architectures for VGG-19 DNN~\cite{qiao2018atomlayer, shafiee2016isaac, song2017pipelayer}. Hence for comparison, we choose VGG-19 network as the representative DNN.
\rev{Moreover, we compare the dynamic power consumption of the DNN hardware since prior work utilizes dynamic power in their results, hence making the comparison consistent.}
The inference latency of the proposed architecture with SRAM arrays is 2.2$\times$ lower than the architecture with ReRAM arrays.
The proposed ReRAM-based architecture achieves 4.7$\times$ improvement in FPS and 6$\times$ improvement in EDAP than AtomLayer~\cite{qiao2018atomlayer}.
The improvement in performance is attributed to the optimal choice of interconnect coupled with the absence of off-chip accesses.
The proposed ReRAM-based architecture consumes 400$\times$ lower power per frame along with 1.74$\times$ improvement in FPS than PipeLayer~\cite{song2017pipelayer}.
Moreover, there is a 5.4$\times$ improvement in inference latency compared to ISAAC~\cite{shafiee2016isaac},
which is achieved by the heterogeneous interconnect structure. 
%
%

\begin{table}[t]
\caption{Inference Performance Results for VGG-19.  
*Reported in~\cite{qiao2018atomlayer}}
\resizebox{.6\textwidth}{!}{
\label{tab:final_result}
\begin{tabular}{@{}ccccc@{}}
\toprule
              & \textbf{Latency} & \begin{tabular}[c]{@{}c@{}}\textbf{Power/frame}\\ \textbf{(W/frame)}\end{tabular} & \textbf{FPS}  & \begin{tabular}[c]{@{}c@{}}\textbf{EDAP}\\ \textbf{(J.ms.mm$^2$)}\end{tabular} \\ \midrule
Proposed-SRAM  & 0.68    & 1.96                                                            & 1458 & 0.46                                                         \\ \midrule
Proposed-ReRAM & 1.49    & 0.43                                                            & 670  & 0.28                                                         \\ \midrule
AtomLayer~\cite{qiao2018atomlayer}      & 6.92    & 4.8                                                             & 145  & 1.58                                                         \\ \midrule
PipeLayer~\cite{song2017pipelayer}      & 2.6*    & 168.6                                                           & 385  & 94.17                                                        \\ \midrule
ISAAC~\cite{shafiee2016isaac}          & 8.0*    & 65.8                                                            & 125  & 359.64                                                       \\ \bottomrule
\end{tabular}
}
\end{table}
\subsection{Connection Density and Hardware Performance}
\begin{figure}[t]
	\centering
	\includegraphics[width=0.7\textwidth]{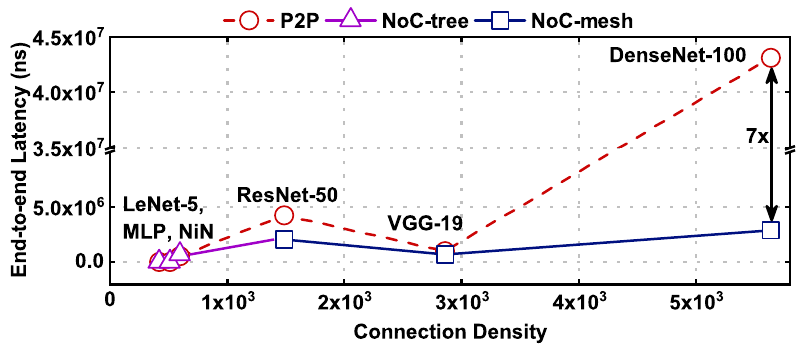}
	\caption{Appropriate selection of NoC topology significantly improves performance for both SRAM- and ReRAM-based IMC architectures.} 
	\label{fig:density_and_latency}
	\vspace{-5mm}
\end{figure}

%
Figure~\ref{fig:ANN} showed a trend of DNNs moving toward a high connection density structure for performance and low connection density structure for compact models. 
Figure~\ref{fig:density_and_latency} shows
the performance for both P2P and NoC-based interconnect at the tile-level for IMC architecture for DNNs with different connection density. We observe a steep increase in total latency with a P2P interconnect.
However, the IMC architecture with NoC interconnect shows a stable curve as we move towards high connection density DNNs.
With the advent of neural architecture search (NAS) techniques~\cite{xie2019exploring, zoph2018learning}, DNNs are moving towards a highly branched structure with very high connection densities. Hence, the NoC-based heterogeneous interconnect architecture provides a scalable and suitable platform for IMC acceleration of DNNs.

%% file: conclusion.tex
\section{Conclusion}\label{sec:conclusion}
The trend of connection density in modern DNNs requires a re-evaluation of the underlying interconnect architecture.
Through a comprehensive evaluation, we demonstrate that the P2P-based interconnect is incapable of handling the high volume of on-chip data movement of DNNs. Further, we provide guidance backed by empirical and analytical results to select the appropriate NoC topology as a function of the connection density and the number of neurons.
We conclude that NoC-mesh is preferred for DNNs with high connection density, while NoC-tree is suitable for DNNs with low connection density.
Finally, we show that the NoC-based heterogeneous interconnect IMC
architecture achieves 6$\times$ lower EDAP
than state-of-the-art ReRAM-based IMC accelerators.


%% file: ack.tex
\section{Acknowledgement}
This work was supported by C-BRIC, one of six centers in JUMP, a Semiconductor Research Corporation (SRC) program sponsored by DARPA, and SRC task 3012.001.